\documentclass[12pt]{iopart}

\DeclareUnicodeCharacter{03B1}{$\alpha$}
\DeclareUnicodeCharacter{03B2}{$\beta$}
\DeclareUnicodeCharacter{03B3}{$\gamma$}

\expandafter\let\csname equation*\endcsname\relax
\expandafter\let\csname endequation*\endcsname\relax

\usepackage{graphicx}
\usepackage{booktabs}
\usepackage{tcolorbox}
\usepackage{placeins}
\usepackage{amsmath}
\usepackage{verbatim}
\usepackage[sorting=none]{biblatex}
\usepackage{enumitem}
\usepackage[separate-uncertainty=true]{siunitx}
\usepackage{fancyhdr}

\usepackage{hyperref}

\usepackage{listings}

\newcommand{\cg}{ChatGPT }
\renewcommand{\theenumi}{\arabic{enumi}}
\bibliography{papers}

\begin{document}
\title{The Impact of AI in Physics Education: A Comprehensive Review from GCSE to University Levels}
\author{Will Yeadon, Tom Hardy}
\address{Department of Physics, Durham University, Lower Mountjoy, South Rd, Durham, DH1 3LE, UK}
\ead{will.yeadon@durham.ac.uk}
\vspace{10pt}
\begin{indented}
\item[]{September 2023}
\end{indented}



\begin{abstract}
With the rapid evolution of Artificial Intelligence (AI), its potential implications for higher education have become a focal point of interest. This study delves into the capabilities of AI in Physics Education and offers actionable AI policy recommendations. Using a Large Language Model (LLM), we assessed its ability to answer 1337 Physics exam questions spanning GCSE, A-Level, and Introductory University curricula. We employed various AI prompting techniques: Zero Shot, In Context Learning, and Confirmatory Checking, which merges Chain of Thought reasoning with Reflection. The AI's proficiency varied across academic levels: it scored an average of 83.4\% on GCSE, 63.8\% on A-Level, and 37.4\% on university-level questions, with an overall average of 59.9\% using the most effective prompting technique. In a separate test, the LLM's accuracy on 5000 mathematical operations was found to decrease as the number of digits increased. Furthermore, when evaluated as a marking tool, the LLM's concordance with human markers averaged at 50.8\%, with notable inaccuracies in marking straightforward questions, like multiple-choice. Given these results, our recommendations underscore caution: while current LLMs can consistently perform well on Physics questions at earlier educational stages, their efficacy diminishes with advanced content and complex calculations. LLM outputs often showcase novel methods not in the syllabus, excessive verbosity, and miscalculations in basic arithmetic. This suggests that at university, there's no substantial threat from LLMs for non-invigilated Physics questions. However, given the LLMs' considerable proficiency in writing Physics essays and coding abilities, non-invigilated examinations of these skills in Physics are highly vulnerable to automated completion by LLMs. This vulnerability also extends to Physics questions pitched at lower academic levels. It is thus recommended that educators be transparent about LLM capabilities with their students, while emphasizing caution against overreliance on their output due to its tendency to sound plausible but be incorrect.
\end{abstract}

\newpage
\begin{section}{Introduction}\label{intro}
\subsection{Background}
\label{subsection-bkgd}
Since OpenAI introduced GPT, there has been a burgeoning interest in the Higher Education (HE) sector regarding the potential impact of Artificial Intelligence (AI) on learning and teaching \cite{prior1, prior2, prior3}. The transformative potential of AI, particularly Large Language Models (LLMs) - neural networks trained on vast amounts of text - has captivated educators. Reinforcing its significance in the educational realm, OpenAI even released a "tips for educators" blog post\footnote{Available at https://openai.com/blog/teaching-with-ai.}. Chatbots like ChatGPT, built on the transformer architecture \cite{transformers}, use a decoder-only design to predict subsequent words, equipping them to handle intricate queries. Following the prominence of ChatGPT, tech behemoths such as Meta, Google, and Baidu have launched their own AI-driven chatbots: LLama, Bard, and Ernie. While these models excel in various tasks, preliminary research indicates that they may not consistently meet the rigorous academic standards of university settings \cite{chatBotComparison}, with GPT-4, the latest iteration of the GPT series, outperforming its counterparts.

Research within Physics specifically has begun to assess the capabilities and implications of AI, largely focusing on ChatGPT. For instance, at the secondary school level, a pilot study led by Bitzenbauer engaged students in asking ChatGPT physics questions as a learning exercise and elicited their feedback on the generated responses \cite{bitzenbauer2023chatgpt}. Moreover, Yeadon and Halliday, when examining a set of Physics exams administered at Durham University, found that GPT-4 typically achieved scores around the 50\% mark \cite{yeadon2023exploring}. Interestingly, the markers frequently noted the plausible sounding nature of the responses from GPT-4, despite them not always being correct. This phenomenon was also highlighted in a study by Dahlkemper et al. \cite{dahlkemper2023physics}. They observed that responses from ChatGPT to challenging Physics questions - ones that were more difficult than the students tested had previously encountered - were rated comparably to researcher-written responses. This was the case even though, for simpler questions, incorrect ChatGPT responses could be identified more easily by the students. Similarly, Gregorcic and Pendrill \cite{gregorcic2023chatgpt} found that a conversation with ChatGPT could yield intricate dialogue but incorrect physics concepts. This issue of complex yet plausible-sounding language masking incorrect content is a hallmark of ChatGPT completions. Focusing on essay-based Physics assignments, researchers discovered that ChatGPT's performance was generally on par with the average student's score on short-form Physics essay assignments \cite{Yeadon_2023}.

A growing body of research suggests the importance of equipping students with skills and experience to interact effectively with AI \cite{bitzenbauer2023chatgpt, kohnke2023chatgpt, adiguzel2023revolutionizing}. While this idea seems practical, it's crucial to acknowledge the continuous evolution of AI and computational technologies. As these systems become more user-friendly, the requirement for extensive technical knowledge decreases. This trend is evident in the rise of intuitive development environments like Replit and design tools like Figma, both of which have simplified many complexities in software creation. A similar trend towards readability is seen in programming languages, with Python being a prime example. Supporting this trend, ChatGPT has shown the capability to convert natural language into functional source code that can solve Leetcode problems \cite{tian2023chatgpt}. Here, natural language can be seen as an even higher-level representation of source code, which itself is a higher-level representation of machine code. This suggests a future where specialized knowledge in areas like prompt engineering could become less important, replaced by more intuitive and direct interaction with AI systems. 

Indeed, the interaction techniques used in this research might soon be outdated due to the rapid development in AI technology. Thus, understanding the effects of different interaction methods on AI performance, along with grasping AI's current capabilities, is vital for the Physics education community. This study aims to delve into these topics to provide educators with a better understanding of how to adapt to the AI evolution and to suggest practical ways to adjust to this rapid change.

\subsection{Interaction with AI models}
\label{subsection-prompt}
There's a growing recognition that the conventional back-and-forth messaging inherent in chat-style interactions may not be entirely representative of the full potential of Large Language Models (LLMs). The interaction quality and the outcome largely depend on not just the proper formulation of the prompt, but also on the application of various frameworks like Reflection and Chain of Thought reasoning. Furthermore, integrating external tools such as Wolfram Alpha can significantly enhance the performance of the LLMs.

Prompting techniques, including Zero Shot and Few Shot, are among the primary factors influencing the output quality. The Zero Shot approach entails asking a question directly and expecting an answer, without providing any prior context or examples. On the other hand, Few Shot involves presenting examples to the model before posing the question \cite{fewShot}, thereby giving the model a context to generate a more informed response. This study uses OpenAI's \cg thus in order to implement Few Shot prompting In Context Learning (ICL) is used whereby the examples are presented within the prompt sent to the LLM.

The Confirmatory Check technique is an implementation that combines elements of the Chain of Thought \cite{CoT} and Reflection \cite{reflection} methods. It encourages a LLM to reconsider its previous outputs, removing excess content if appropriate. This method prompts the model to evaluate its initial response, thereby mitigating the problem where the LLM becomes 'stuck' with a mistake in its produced answer. Additionally, LLM outputs can sometimes be long, rambling, and inconsistent with the complexity of the question. The Confirmatory Check technique provides an opportunity for the LLM to avoid these issues. This approach offers valuable insights in educational contexts, where it mimics a more conversational interaction between students and the LLM. This back-and-forth dialogue can lead to more refined and accurate answers.

The idea of equipping LLMs with external tools to handle challenging tasks has also gained traction recently. A notable instance is the integration of Wolfram Alpha with ChatGPT, allowing the LLM to leverage Wolfram Alpha's capabilities to tackle complex mathematical tasks that are typically difficult for LLMs \cite{chatGPTmathCapabilities}.

In exploring these different techniques, our aim is not only to provide a broader understanding of how these models can be utilized but also to evaluate their efficacy within the context of Physics education. This serves the larger goal of this study - to benchmark these cutting-edge LLMs thoroughly and provide educators with a clearer picture of AI's strengths and weaknesses. By doing so, we hope to offer a comprehensive resource to understand AI's current capabilities and thereby inform educators about effective ways to integrate AI into their teaching practices.
\end{section}

\begin{section}{Method}
\subsection{Question sources}
To ensure a comprehensive examination of the AI's capabilities across various difficulty levels, we sourced questions that spanned from GCSE to A-Level, as well as introductory university courses. These questions were obtained from a wide array of educational boards and institutions, culminating in a diverse and robust dataset. To transfer questions from their original sources into a digital, machine-readable format, we utilized a combination of regular expression matching and manual transcription. Special emphasis was placed on maintaining the accuracy of the transcription process, preserving the original complexity and structure of each question. However, due to the requirement of sending API requests in Latin-1 encoding (ISO/IEC 8859-1), mathematical notations such as the square root or integral symbols were unavailable. We adapted to this constraint by using natural language short-hands, such as 'sqrt(x)' or 'integrate(x)', which proved to be an effective solution. Further, when questions incorporated tables or figures, we adopted specific strategies. Tables were reformatted to resemble nested Python lists. As for figures, we provided detailed descriptions. However, this approach for figures was seldom practical. The questions were organized into three distinct categories: Numerical, where calculations such as "find the acceleration" were required; Multiple Choice, involving selection from a list of options; and Written Descriptive Answers, where textual responses were needed. The sources of the questions are detailed in Table \ref{table:source}.

\begin{table}[ht]
\centering
\begin{tabular}{lp{9cm}p{2cm}}
\toprule
Level & Source & \begin{tabular}[c]{@{}l@{}}Number of \\ Questions\end{tabular} \\
\midrule
GCSE & OCR Physics A - Gateway Physics 2017-21 & 263 \\
& OCR Physics B - 21st Century Physics 2017-21 & 91 \\
A-Level & OCR Physics A 2017-21 & 244 \\ 
& OCR Physics B (Advancing Physics) 2017-21 & 130 \\
University & University Physics with Modern Physics 15th Ed. & 175 \\
& Physics for scientists and engineers 4th Ed. & 180 \\
& College Physics 2e 2nd Ed. & 112 \\
& Physics Principles With Applications 7th Ed. & 142 \\
\midrule
Total & & 1337 \\
\bottomrule
\end{tabular}
\caption{Question sources used for the evaluation.}
\label{table:source}
\end{table}


Building on the extensive research focusing on university-level physics exam questions~\cite{yeadon2023exploring}, the current study narrows its scope to introductory level questions. The textbooks from which these questions were sourced are shown in Table~\ref{table:source}. To ensure fair evaluation, the scoring procedures for these questions were standardized across the different educational levels. For GCSE and A-Level questions, we adhered to the respective mark schemes provided. On the other hand, university-level questions, being derived from textbooks, lacked a standard mark scheme. To address this, a customized scoring rubric was developed. Specifically, questions from \textit{University Physics with Modern Physics} were found to be more elaborate and were consequently marked on a 2-point scale. A score of 2/2 was awarded for completely accurate answers, 1/2 for answers that were near correct with correct application of physics principles, and 0/2 for all other responses. For questions sourced from the other university textbooks, a simpler 1-point scale was utilized, wherein each question was marked as either correct or incorrect. This approach aimed to strike a balance between accommodating the inherent complexity of questions from different sources and maintaining an equitable evaluation framework.

\subsection{Generating the AI answers}
\label{subsection-sample-gen}
We utilized the OpenAI API, specifically the GPT-3.5-turbo language model, to generate AI responses from an array of message objects \cite{openai_api}. By altering the format of the message object array, we implemented various interaction techniques: Zero Shot, In Context Learning and Confirmatory Check. Each message object has a role of either \emph{system}, \emph{user}, or \emph{assistant}. The \emph{system} message objects guide the behavior of the LLM. The OpenAI default \emph{system} message is 'You are a knowledgeable assistant.' \cite{openai_best}, which was retained for the Zero-shot prompting interactions. It was followed by a \emph{system} message reading 'Please answer the following question.' to ensure the question was answered, and then a \emph{user} message containing the actual question content.

For the In Context Learning prompt implementation, the system messages were modified to include a series of example question-answer pairs before the target question, as shown in Figure \ref{fig-fewShot}. These examples served to establish the context for the expected responses. Studies have shown that beyond five examples, the benefits of additional examples become negligible \cite{dong2022survey}. We found the LLM would often provide lengthy responses, so the examples were deliberately concise. Although a word-based example was initially included, it was determined to be unnecessary since the model is trained principally on long text passages.

\begin{figure}[htp]
    \centering
    \begin{lstlisting}[basicstyle=\ttfamily\small, breaklines=true, frame=single]
[{"role": "user", "content": """
     A 30W light bulb uses 600J of electrical energy in time t to produce 450 J of light energy. What is the efficiency eta of the light bulb?
    """ },
    {"role": "assistant", "content": """
     The efficiency is the ratio of useful energy output to the total energy input expressed as a percentage so in this case: 
         eta = (450 J / 600 J) * 100%
         eta = 75 %
     """ },
    {"role": "user", "content": """
     Interference fringes, produced by monochromatic light are viewed on a screen placed a distance L from a double slit system with slit separation S. The distance between the centres of two adjacent fringes (the fringe separation) is W. If both S and L are doubled, what will be the new fringe separation?
         A) 2W
         B) W/2     
         C) W
         D) 4W
    """ },
    {"role": "assistant", "content": """
    C
     """ },
    {"role": "user", "content": """
     A car accelerates from 12 m s to 21 m s in 6.0 s. How far did it travel in this time? Assume constant acceleration.
    """ },
    {"role": "assistant", "content": """
    First, we'll find the acceleration using the equation: a = (v - u) / t = (21 - 12) / 6 = 1.5 m/s^2

    Next, we'll use the equation for the distance traveled: s = ut + (1/2) * a * t^2 = 12 * 6 + (1/2) * 1.5 * 36 = 99 m

    So, the car traveled a distance of 99 meters during this time.     
     """ },]
    \end{lstlisting}
    \caption{Message array used to implement the Few-shot prompting via In Context Learning, illustrating how context is provided to guide the Language Model's responses.}
    \label{fig-fewShot}
\end{figure}

The Confirmatory Check technique was implemented by sending the In Context Learning message object to the API with the In Context Learning response appended as an \emph{assistant} message. It was followed by a \emph{user} message reading, \emph{"Please check the previous answer to ensure you're happy with it. If you feel that you can express it more succinctly, then please do so. For reference, this was the original question: $<$question inserted$>$.}" This approach allowed the LLM an opportunity to refine its In Context Learning answer.

While the OpenAI API does not directly provide a confidence score or probability with each response, the 'temperature' parameter was set at 0 to eliminate randomness in the generated responses \cite{openai_temperature}. The 'max tokens' parameter was set at 2000, suitable for extensive answers. After processing each question, the result was saved with the new answers in an Excel workbook to prevent data loss in case of program termination. The grading and interpretation of the AI's responses are discussed in the subsequent sections.

\subsection{Automated grading}
\label{subsection-grading}
To assess the LLM's capability in evaluating its own responses, the answer from each question - spanning different prompting styles - was submitted to the API. This submission included the solution, marking guidance, available marks, and the original question\footnote{A overview of the prompt instructing the AI to mark its own work can be found in the Appendix Figure \ref{fig-marking}.}. Tasking the LLM with marking its answers emulates a human marker's role. Subsequent comparisons between LLM-assigned scores and human evaluations provided insights into the LLM's efficacy. Given the potential for the LLM to assign improbable scores, like values below zero or exceeding available marks, checks were put in place. If an invalid score was provided thrice consecutively, it was recorded as '-1' signifying a marking failure. Due to the comprehensive marking guidance availability, only GCSE and A-Level sources were utilized.

\subsection{Mathematical capabilities}
\label{subsection-math-capabilities}
To assess the mathematical abilities of the LLM, two datasets comprising a total of 5,000 numbers were generated. The first dataset contained 2,500 integer pairs stratified based on the number of digits, ranging from 1-digit to 5-digit numbers. Within each stratum of 500 pairs, a random arithmetic operation (addition, subtraction, multiplication, or division) was assigned, ensuring an approximate distribution of 125 operations per digit-length category. The second dataset focused on single operand operations applied to 2,500 integers. Operations included squaring, square-rooting, calculating the natural logarithm, \(\sin\) or \(\cos\). Each digit-length category in this dataset had approximately 100 of each operation. The primary objective was to gauge the LLM's computational accuracy, especially in relation to numerical complexity. Responses were assessed based on perfect match accuracy, deviation within 5\% of the correct answer, and deviation within 10\% of the correct answer. For the purposes of determining a 'perfect match', responses were considered correct if they were accurate up to 5 decimal places. Detailed insights derived from these analyses are delineated in Section~\ref{subsection-math-results}.
\end{section}

\begin{section}{Analysis and results}
\label{section-analysis}
\subsection{Overview}
\label{subsec-analysis-overview}
\begin{figure}[!htbp]
\centering
\includegraphics[width=15cm]{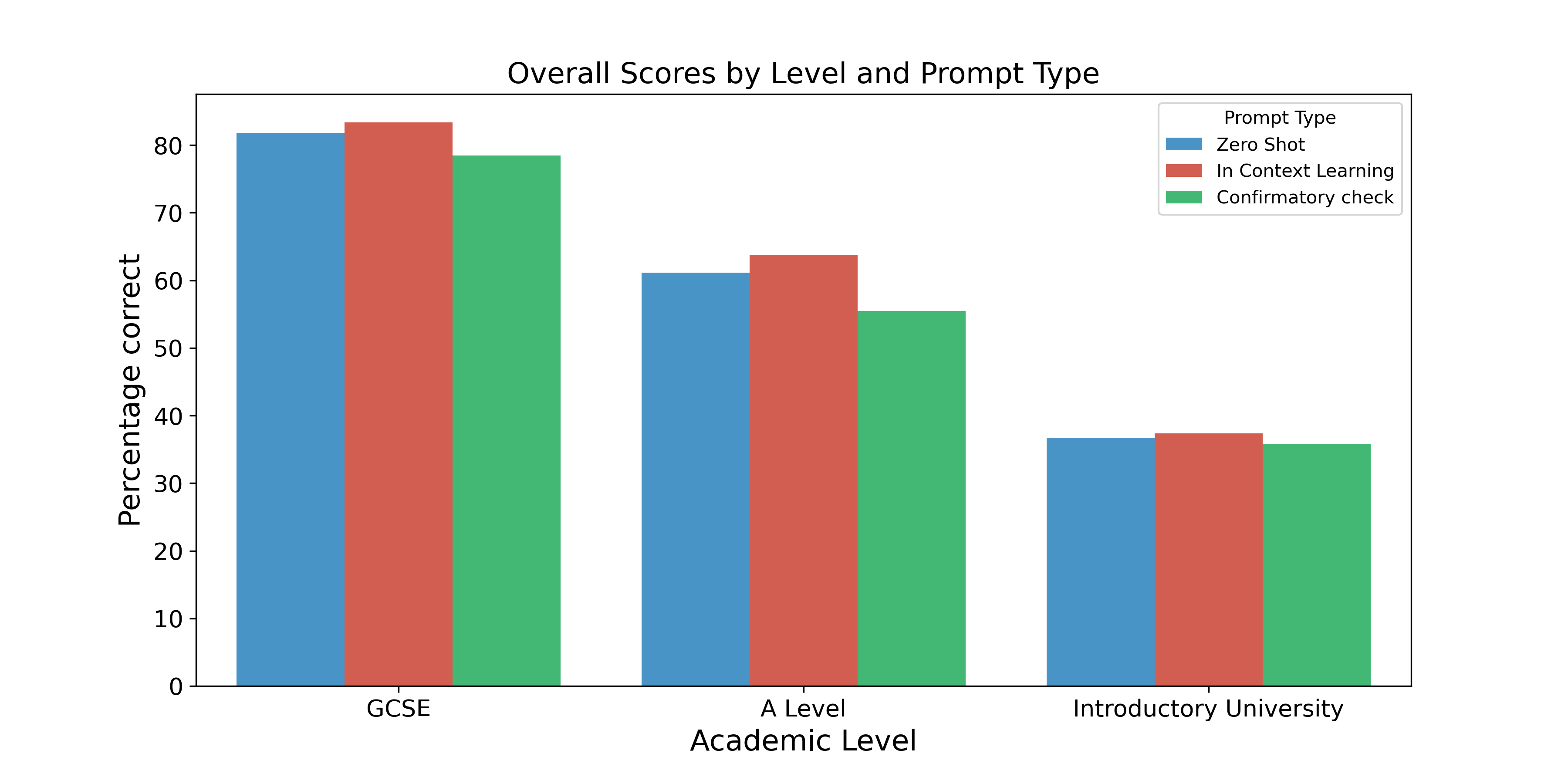}
\caption{Comparative analysis of overall scores achieved by different AI prompting techniques (Zero Shot, Few Shot, Confirmatory Check) across three academic levels: GCSE, A Level, and Introductory University.}
\label{fig-prompt}
\end{figure}

\begin{figure}[!htbp]
\centering
\includegraphics[width=15cm]{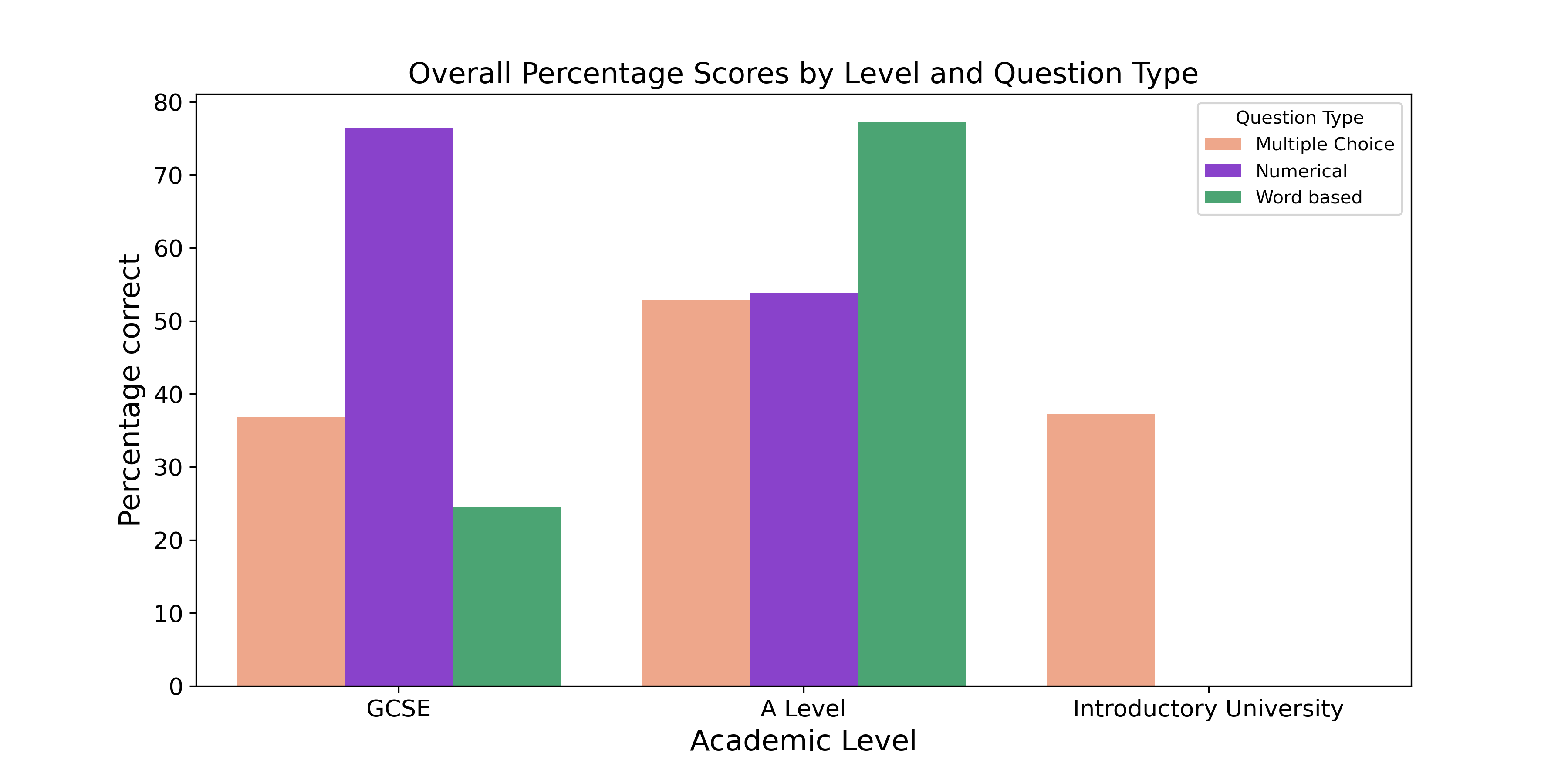}
\caption{A detailed breakdown of the AI's performance in terms of percentage correct for various question types (Multiple Choice, Numerical, Word-based) at different academic levels, highlighting areas of success and potential improvement.}
\label{fig-question}
\end{figure}

Figure \ref{fig-prompt} illustrates the overall scores achieved by different AI prompting techniques across three academic levels: GCSE, A Level, and Introductory University. The three techniques represented are Zero Shot (blue), In Context Learning (red), and Confirmatory Check (green). It shows that the performance of the three prompting techniques remains relatively consistent across the three academic levels, while the overall performance decreases as the academic level increases. Although there are slight variations in the percentage of correct answers, none of the techniques consistently outperforms the others across all levels.

To determine if these observed differences were statistically significant, an Analysis of Variance (ANOVA) test was conducted, with the results summarized in Table \ref{tab:anova_results}. ANOVA is particularly apt for this analysis as it allows for a comparison of means across more than two groups. The null hypothesis for the ANOVA test states that there is no significant difference between the group means. The alternative hypothesis posits that at least one group mean is different. For the GCSE, A Level, and Introductory University levels, the p-values were 0.5429, 0.1310, and 0.8828, respectively, indicating that we fail to reject the null hypothesis for all three academic levels. This suggests that the choice of prompting technique does not play a pivotal role in the AI's performance.

For a more nuanced analysis, each question was categorized as either Multiple Choice, Numerical, or Word-based. However, at the Introductory University level, the dataset is overwhelmingly composed of numerical questions ($>99\%$). This dominance renders a detailed, segregated analysis by question type challenging for this academic level. Nevertheless, the ANOVA test results for the GCSE and A Level, as showcased in Table \ref{tab:anova_results}, indicate a statistically significant difference in the performance of the three prompting techniques across the various question types. Yet the differences are not consistent between academic levels with the LLM performing best on numerical questions at GCSE but best on word based at A-Level. Further word based questions were the worse performing type for the LLM at GCSE. The nature of the question can notably affect the LLM's accuracy. For example, in multiple choice questions, the LLM frequently settled on an answer that wasn't among the provided options. In these scenarios, it either refrained from answering altogether or selected the option that was closest to its often incorrect answer. Beyond these question types, it was observed that questions with tables scored similarly to those without, indicating that tables do not hinder LLM performance.


\begin{table}[ht]
\centering
\caption{ANOVA Results for Different Prompting Techniques and Question Types}
\label{tab:anova_results}
\begin{tabular}{l|c|c}
\hline
\textbf{Category} & \textbf{F-statistic} & \textbf{p-value} \\
\hline
\multicolumn{3}{c}{\textbf{Overall ANOVA Results}} \\
\hline
GCSE Level & 0.6111 & 0.5429 \\
A Level & 2.0366 & 0.1310 \\
Introductory University Level & 0.1246 & 0.8828 \\
\hline
\multicolumn{3}{c}{\textbf{GCSE Level ANOVA Results for Different Question Types}} \\
\hline
Zero Shot & 80.7864 & \(1.3781 \times 10^{-29}\) \\
In Context Learning & 88.5268 & \(7.4370 \times 10^{-32}\) \\
Confirmatory Check & 92.3540 & \(5.9499 \times 10^{-33}\) \\
\hline
\multicolumn{3}{c}{\textbf{A Level ANOVA Results for Different Question Types}} \\
\hline
Zero Shot & 46.5909 & \(8.8761 \times 10^{-19}\) \\
In Context Learning & 57.8985 & \(1.3057 \times 10^{-22}\) \\
Confirmatory Check & 41.2386 & \(6.7268 \times 10^{-17}\) \\
\hline
\end{tabular}
\end{table}

\subsection{Example question answer}
\label{subsec-analysis-illustrative}

Looking at specific examples offers a clear perspective on the influence of prompt engineering. As depicted in Figure \ref{promptComparison}, the nuances of different prompting styles can lead to varied responses. Given the Physics question '\emph{Write a decay equation in terms of a quark model for beta-minus decay}' the Zero-Shot prompt failed to appreciate the question asked about the quark model instead detailing $\beta ^-$ in a nucleus. The In Context Learning prompting got the question completely correct but Confirmatory Check approach lost marks due to it stating an electron neutrino rather than an electron antineutrino in the answer. Interesting this may have been because a actual $\nu$ character was returned instead of the words 'anti-v' but $\bar{\nu}$ isn't available in the Latin-1 character set.

\begin{figure}[htp]
    \centering
    \begin{tabular}{|l|p{10cm}|}
    \hline
    \textbf{Prompt Type} & \textbf{Response} \\
    \hline
    Zero-shot & In the quark model, beta-minus decay can be represented by the following decay equation: \\
              & n $\rightarrow$ p + e- + anti(v)e \\
              & where "n" represents a neutron, "p" represents a proton, "e-" represents an electron, and "anti(v)e" represents an antineutrino. \\
    \hline
    In Context Learning & In terms of a quark model, beta-minus decay can be represented by the following decay equation: \\
                        & d $\rightarrow$ u + e- + anti-v(e) \\
                        & In this equation, the down quark (d) decays into an up quark (u), an electron (e-), and an electron antineutrino (anti-v(e)). \\      
    \hline
    Confirmatory Check & In terms of a quark model, beta-minus decay can be represented by the following decay equation: \\
                       & d $\rightarrow$ u + e- + $\nu$e \\
    \hline
    \end{tabular}
    \caption{Comparison of responses for the given question based on different prompting styles in response to the question '\emph{Write a decay equation in terms of a quark model for beta-minus decay}'.}
    \label{promptComparison}
\end{figure}

The Zero-shot approach, while thorough, often yielded verbose answers, averaging 427 characters in length. In contrast, the In-context Learning method trimmed responses to an average of 405 characters. The Confirmatory Check approach stood out as the most concise, with answers averaging just 228 characters. Additionally, while some mathematical content in the responses mirrored conventional formats, there were instances where the representations, though appearing correct, were mathematically inaccurate.

\subsection{AI marking}
\label{subsec-analysis-AI-grading}
For this evaluation, only instances where both human and the LLM successfully assigned a grade were included. Out of 3486 AI-generated answers to 1162 questions\footnote{Questions from \emph{University Physics with Modern Physics} had a separate solutions booklet so couldn't be marked by the AI}, the LLM only successfully graded 2209 instances, achieving a 63.4\% rate of successful evaluations. All scores were normalized to facilitate a fair comparison across questions with different maximum marks. Human and LLM evaluation showed a concordance in scores for Zero-shot, In Context Learning, and Confirmatory Checking with rates of 49.82\%, 51.96\%, and 50.54\%, respectively. This means that for approximately half of the questions, the LLM gave the same score as the human markers. Among these 2209 graded instances, human markers assigned an average normalized score of 0.515, with a standard deviation of 0.448. The LLM's average normalized score was a lot higher at 0.952 but had a lower standard deviation of 0.167.

\begin{table}[h]
\centering
\caption{Correlation Matrix for grades assigned by Humans and the LLM. Human-ZS: Zero-shot prompted answers; Human-ICL: In Context Learning prompted answers; Human-CC: Confirmatory Check prompted answers; LLM-ZS, LLM-ICL, and LLM-CC are analogous for the LLM.}
\label{corr-mat}
\vspace{1.0ex} 
\begin{tabular}{lcccccc}
\toprule
          & Human-ZS & Human-ICL & Human-CC & LLM-ZS & LLM-ICL & LLM-CC \\
\midrule
Human-ZS     & 1.000 & 0.800 & 0.754 & 0.189 & 0.156 & 0.139 \\
Human-ICL    & 0.800 & 1.000 & 0.913 & 0.110 & 0.241 & 0.201 \\
Human-CC     & 0.754 & 0.913 & 1.000 & 0.102 & 0.197 & 0.257 \\
AI-ZS        & 0.189 & 0.110 & 0.102 & 1.000 & 0.334 & 0.284 \\
AI-ICL       & 0.156 & 0.241 & 0.197 & 0.334 & 1.000 & 0.662 \\
AI-CC        & 0.139 & 0.201 & 0.257 & 0.284 & 0.662 & 1.000 \\
\bottomrule
\end{tabular}
\end{table}

The observed correlations in Table \ref{corr-mat} show that human markers often grade In Context Learning and Confirmatory Checks in a correlated manner, evidenced by a strong internal correlation of 0.913. In contrast, AI markers displayed a slightly weaker internal correlation of 0.662 between these same methods. Comparing human and AI grading reveals a moderate level of agreement, particularly for In Context Learning (ICL) and Confirmatory Checking (CC) with correlation values of 0.241 and 0.257, respectively. Zero-shot prompting shows a weaker correlation of 0.189.

Understanding these correlation values alongside the concordance rates suggests that the agreement is higher for straightforward questions with single correct answers. Meanwhile, more complex questions are likely sources of disagreement. These discrepancies may arise from the LLM's different interpretation of the marking guidance or its emphasis on different parts of the response. The LLM-assigned scores also have a lower standard deviation, indicating a more consistent but potentially less nuanced grading approach.

%

\subsection{Mathematical capabilities}
\label{subsection-math-results}
For the two-integer operations, the LLM achieved an exact accuracy rate of 52.3\% across all questions. When a tolerance of $\pm 5\%$ of the exact answer was considered, the accuracy rate climbed to 75.8\%. The analysis uncovered significant variances in performance depending on the arithmetic operation involved. While the model demonstrated high accuracy in addition and subtraction across all levels of numerical complexity, its performance in multiplication and division was less reliable, especially with higher-digit numbers. Figure~\ref{fig-maths} illustrates these findings, highlighting how the LLM's accuracy is influenced by both the operation type and the numerical complexity involved. These results suggest caution when employing the LLM for tasks requiring precise numerical calculations, as its performance can be operation and complexity dependent.

\begin{figure}[!htp]
    \centering
    \includegraphics[width=13.5cm]{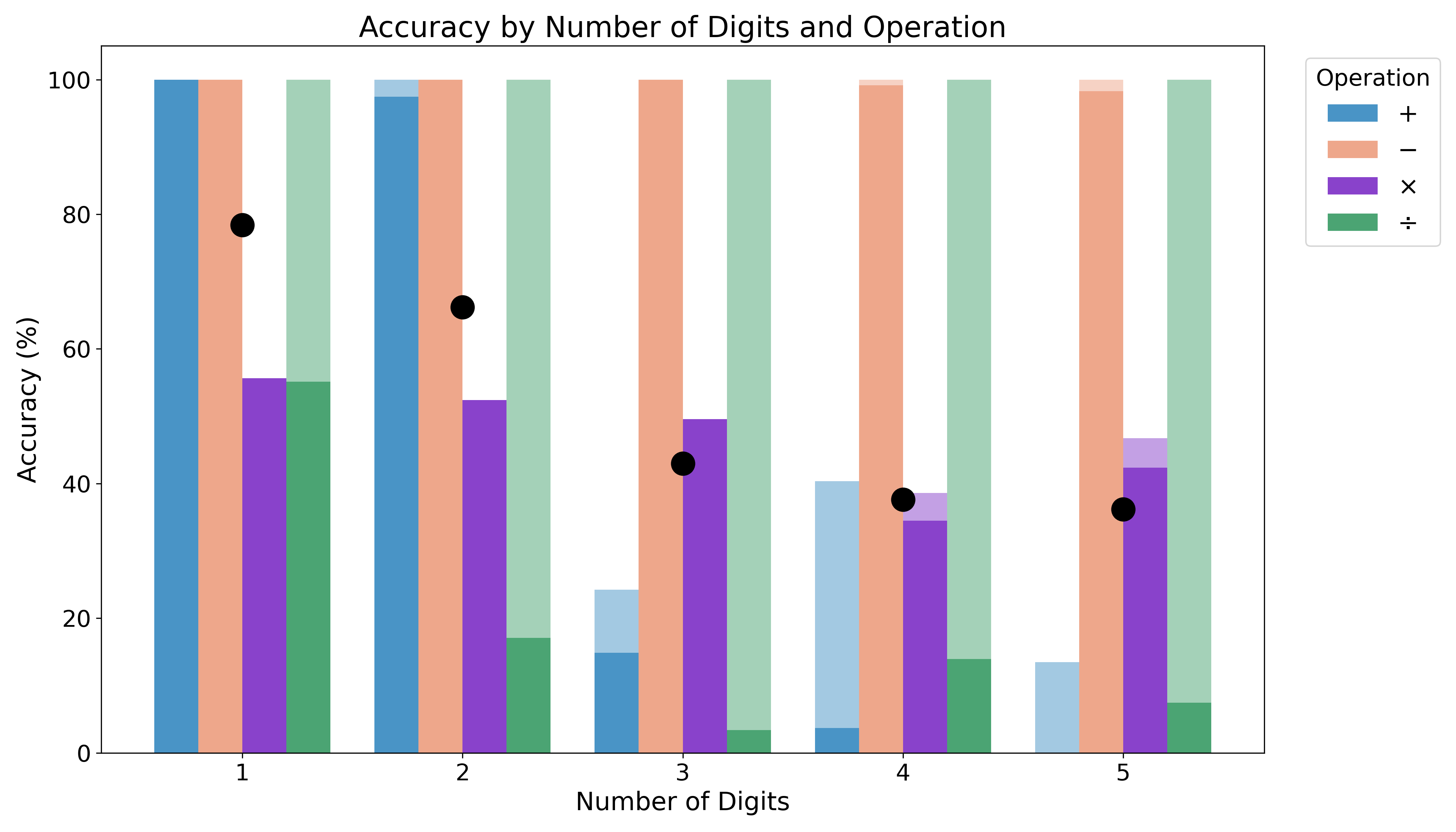}
    \caption{Accuracy of the LLM in performing basic arithmetic operations across varying numerical complexity. The bars represent exact accuracy, while the transparent overlay indicates accuracy within a 5\% margin. The black dot markers denote the overall average accuracy for each digit length, across all operations.}
    \label{fig-maths}
\end{figure}

\begin{figure}[!htp]
    \centering
    \includegraphics[width=13.5cm]{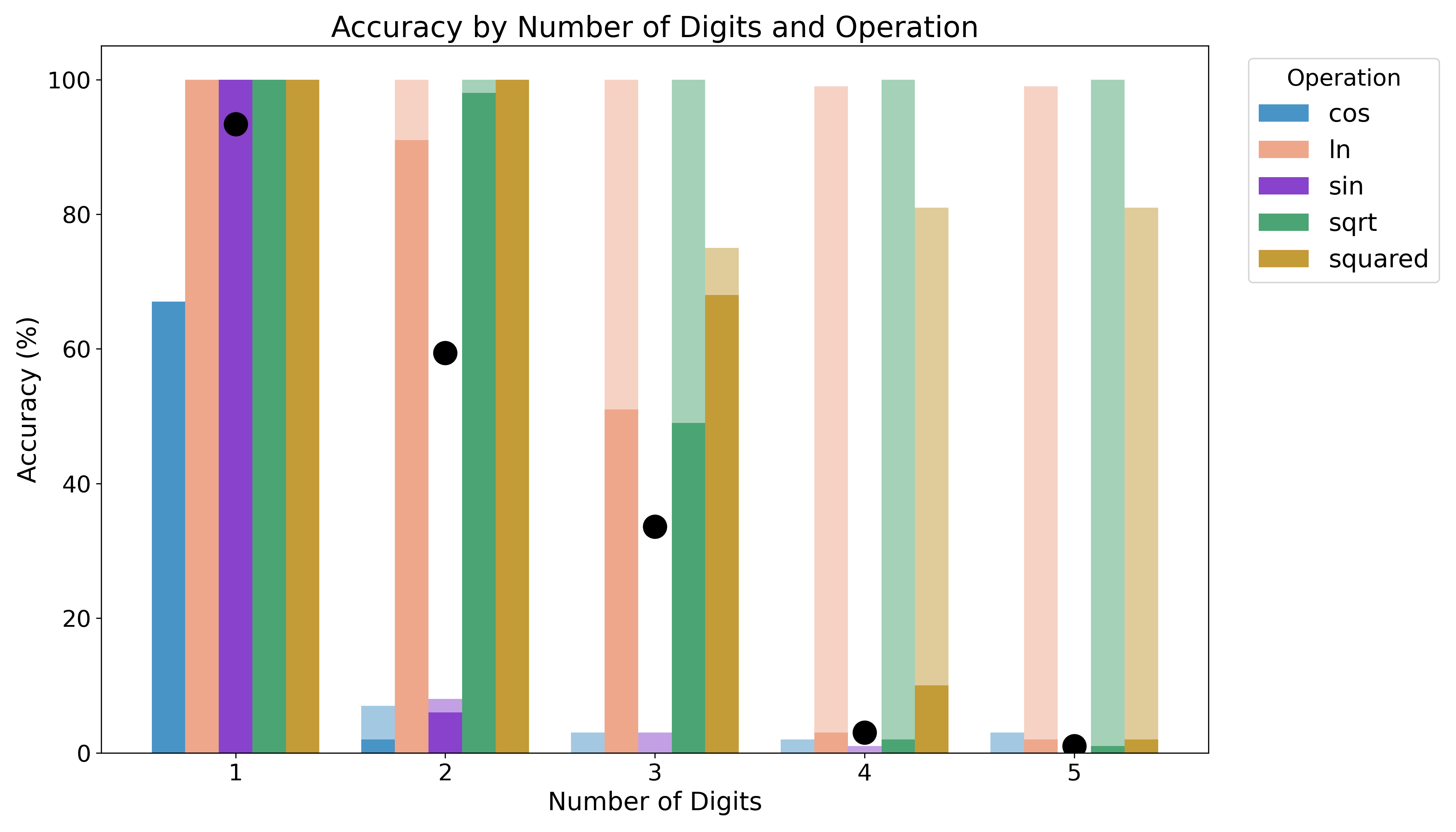}
    \caption{Accuracy of the LLM in performing common mathematical functions across varying numerical complexity. The bars represent exact accuracy, while the transparent overlay indicates accuracy within a 5\% margin. The black dot markers denote the overall average accuracy for each digit length, across all functions.}
    \label{fig-maths-p2}
\end{figure}

When evaluating operations involving single operands, the LLM achieved an exact accuracy rate of 38.1\% on complex mathematical functions. Introducing a margin of error displayed some improvement: the accuracy ascended to 63.2\% with a 5\% tolerance. Delving deeper into the individual functions, it was evident that the LLM struggled with trigonometric functions when handling more than a single digit. Contrastingly, operations like the natural logarithm and squaring maintained commendable performance - even with larger numbers, they mostly stayed within a 5\% tolerance. Figure~\ref{fig-maths-p2} visually underscores these insights, showing a general trend where performance diminishes with an increase in the number of digits. This observed trend mirrors the pattern from basic arithmetic operations, reinforcing the notion that the LLM's capability diminishes with heightened numerical complexity.
\end{section}

\begin{section}{Discussion and Conclusion}
\label{section-disc}
\subsection{Overview}
Artificial Intelligence, especially in the realm of Large Language Models (LLMs), continues to draw attention in academic circles. Within this landscape, this study set out to evaluate the proficiency of AI in Physics Education. The results presented in this study and elsewhere allow us to make general conclusions about LLM use within Physics Education and to provide recommendations for educators. 

For the characteristics of LLM output, one notable aspect is that without a specific syllabus to adhere to, LLMs often introduced innovative methods, leading to novel approaches in answering. While this can be a fresh perspective, it does not always align with the traditional academic evaluations. Contrary to prior work emphasizing the importance of good prompting \cite{promptEngineer1, promptEngineer2}, our investigation revealed statistically insignificant difference between different interaction techniques. We found that AI struggles with harder Physics, as shown in Figure~\ref{fig-question}. As the academic level increased, the amount of correct responses decreased. Previous research has highlighted how AI can often struggle with more complex Physics; beyond introductory textbooks, Yeadon et al. \cite{yeadon2023exploring} demonstrated how GPT-3.5 typically fails to pass most Physics exams at Durham University. However the latest foundation model GPT-4 consistently outperforms GPT-3.5 and often scores nearly 50\% on exams, this is shown in Figure~\ref{fig-all}. Given these results, and as highlighted at the end of \cite{chatBotComparison}, the current potential threat of AI in non-invigilated online exams at university level seems to be relatively contained. In fact, it would be prudent to warn students that AI performance at GCSE and A Level may not transfer to university assessments. This leads to the conclusion that whilst non-invigilated GCSE and A-Level assessments should be wary of how good the latest foundational AI models are, at University level the threat is not as dire. The score of the best AI systems seems to, on average, peak at around 50\% for Physics questions meaning currently only the weaker students would benefit.

\begin{figure}[!htp]
\centering
\includegraphics[width=13.5cm]{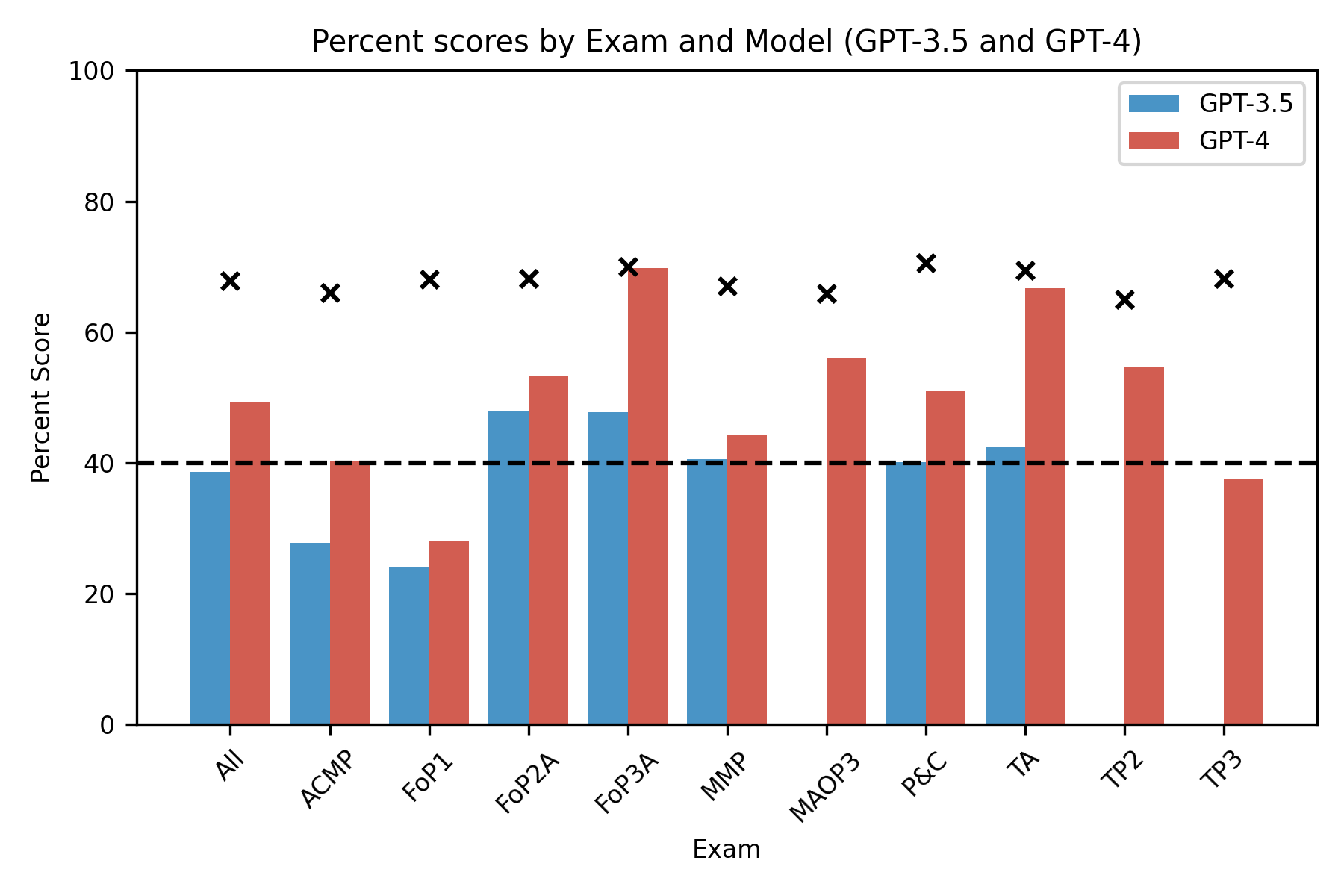}
\caption[Performance of GPT-4 and GPT-3.5 on Physics exams]{Performance of GPT-4 and GPT-3.5 on different physics exams as presented by Yeadon et al. \cite{yeadon2023exploring}. The black crosses indicate the average student mark from 2018 - 2021 on the modules for the exam and the dashed black line shows the 40\% score required to pass the exam. Critically, these exams were marked by the same academics who mark student exams. Acronym definitions are provided below.\protect\footnotemark Adapted with permission from \cite{yeadon2023exploring}.}
\label{fig-all}
\end{figure}
\footnotetext{The modules acronyms are for: 
ACMP : Advanced Condensed Matter Physics,
FoP1 : Foundations of Physics 1,
FoP2A : Foundations of Physics 2A,
FoP3A : Foundations of Physics 3A,
MMP : Mathematical Methods in Physics,
MAOP3 : Modern Atomic and Optical Physics 3,
P\&C : Planets and Cosmology,
TA : Theoretical Astrophysics,
TP2: Theoretical Physics 2,
TP3: Theoretical Physics 2.
}

As a part of a Physics degree, often there are written elements and computational work. Here the threat to assessment fidelity is more pronounced. There are LLMs specifically trained on coding examples which can excel at complex coding tasks found in a computer science focused degree where the complexity would typically be beyond that found in a Physics degree \cite{codePerform1, codePerform2}. Further, research looking at Physics essays specifically found AI excels here \cite{Yeadon_2023}. It is important that educators are aware of the capabilities in these areas and it is recommended that for coding and essay work, if the assessment is non-invigilated educators should enter their assignments into GPT-4 and see the capabilities themselves. The wide availability and capability of modern LLMs may be irreconcilable with with take home short essays or typical Physics coding tasks.   

LLMs often produce verbose outputs, the AI's proclivity to produce extensive responses, often not proportional to the question's complexity, is not only a hallmark but seems to be an integral part of quality answers. Whilst not statistically significant, there was a decline in performance with the Confirmatory Checking raises concerns about the AI's current capacity for iterative, conversational interaction, resonating with the observations by \cite{gregorcic2023chatgpt}. Interestingly, looking at the linguistics of the output much prior research has highlight how AI generated content is both difficult to detect \cite{detect1, detect2} and potentially bias against non-native English speakers \cite{gptBiasEnglish}. Curiously there are simple techniques to get the AI to reveal itself such as asking 'Do you agree with this statement?' will often get the LLM to state 'As an AI assistant I do not have personal opinions, emotions, or preferences.'. Similarly the use of zero-width spaces or hidden prompt injection attacks \cite{injection} within questions can also foil LLM effectiveness.

\begin{figure}[ht]
    \centering
    \begin{tcolorbox}[
    colframe=black!25,
    colback=yellow!10]
    \begin{enumerate}
         \item \textbf{Novel Approaches:} LLMs, unbound by a specific syllabus, frequently adopt innovative methods when addressing questions. Such distinct approaches can serve as an indication that LLMs might have been employed in crafting the answer.
         
         \item \textbf{Mathematical Hurdles:} Pure LLMs, lacking computational tools, face challenges when handling mathematical operations, particularly with numbers exceeding three digits.
         
         \item \textbf{Prompting Limitations:} Contrary to initial beliefs about best-practice prompt engineering, this study revealed its apparent limited efficacy in the realm of Physics questions. While ICL enhancements did yield improved results, the advantage over Zero-shot approaches was marginal.
         
         \item \textbf{Verbose Outputs:} The AI consistently produces verbose answers, often misaligned with the question's complexity. Notably, when provided an opportunity to refine its outputs, the LLM frequently produced content of diminished quality.
         
        \item \textbf{Graphical Challenges:} 
        In this study, an effective method for LLMs to address graphical questions was not identified. With minimal reformatting, LLMs handled questions involving tabled data comparably to other queries. Moreover, for 'sketch' / 'diagram' tasks, LLMs frequently used rows of symbols, offering reasonable attempts.
 
    \end{enumerate}
    \end{tcolorbox}
    \caption{Key conclusions derived from this study's assessment of LLM responses to Physics questions.}
    \label{fig-conc-updated}
\end{figure}

The present work also highlighted how LLMs can struggle with mathematical computations as the lengths of digits involved increases. Of the 5000 mathematical questions asked only 45.2\% were answered correctly. The difference here however is that modern computers already have sophisicated mathematical capabilities meaning it would be inapt to use a LLM to work out the cosine of a number when calculators are available. The AI's grading capability further supports this viewpoint, when marking multiple choice questions the AI often struggled to do this simple task correctly, a case of over engineering / using the AI for the wrong task. In fact when extending the marking to all questions a congruence rate of only 50.8\% with human evaluations was found, indicating clear limitations in certain areas.

On a positive note, during our interactions, the AI maintained a respectful tone without displaying any abusive or exclusionary language, reflecting advancements in ethical AI design. While premium versions of some technologies might be inaccessible to some due to cost, educators should ensure that no student is mandated to use paid resources. To summarize, while AI has made significant strides, limitations persist in its application to Physics. The key conclusions from our study are outlined in Figure~\ref{fig-conc-updated}.

\subsection{Recommendations}
The swift progress in AI technology raises numerous ethical dilemmas, especially regarding its potential misuse in academia, its inherent biases, and its overarching societal repercussions. Echoing the concerns raised by \cite{cooper2023examining}, the incorporation of AI into the educational realm warrants a balanced mix of skepticism and meticulous scrutiny. As AI models continuously advance, a shared responsibility falls upon educators, developers, and policymakers to maintain vigilance, ensuring that AI tools are harnessed ethically and judiciously. In light of the current state of affairs, specific recommendations are posited, as depicted in Figure~\ref{fig-rec-updated}. 

\begin{figure}[ht]
    \centering
    \begin{tcolorbox}[
    colframe=black!25,
    colback=blue!10]
    \begin{enumerate}
         \item \textbf{Transparency About Capabilities:} With over 100 million users, ChatGPT's influence is undeniable, and it's frequently highlighted in the news. Educators should openly discuss its strengths and weaknesses with students, especially its propensity to produce plausible yet occasionally incorrect or incomplete answers. 
         
         \item \textbf{Caution Students Against Overreliance:} While AI may prove valuable at GCSE levels, its effectiveness can diminish in university settings, as illustrated in Figure~\ref{fig-prompt}. Students should be reminded that relying heavily on AI can deprive them of genuine learning experiences.
         
         \item \textbf{Avoid Teaching AI Interaction Techniques:} The study found no significant variance in performance across different prompt engineering methods for Physics questions. This was surprising as effective prompting techniques are subject of much research and reported improved performance \cite{SC-CoT, ZS-CoT, CoT}. However, from a Physics teaching perspective there is not enough clear benefit in improving Physics question answering abilities. Further, given the rapid advancements in AI, previously effective techniques can soon become outdated.

        \item \textbf{Change Some Assessment Methods:} Non-invigilated coding and short form essays are very vulnerable to automated completion by LLMs \cite{Yeadon_2023}. Further, as AI-written text is difficult to discern \cite{detect1, detect2} and potentially bias against non-native English speakers \cite{gptBiasEnglish}. Advertised AI detectors should not be trusted.  

        \item \textbf{Anticipation of Evolving Capabilities:} Educators should stay updated with the latest in AI advancements. As Yeadon and Halliday's study \cite{yeadon2023exploring} illustrates, there's a discernible improvement from GPT-3.5 to GPT-4. However, it remains uncertain whether future models will improve further still or approach an asymptote.

        \item \textbf{Ethical Considerations in AI Use:} AI's interaction has shown a respectful tone without exclusionary language, highlighting advancements in ethical AI design. However, educators should ensure equitable access by not mandating the use of premium, potentially inaccessible technologies for students.

    \end{enumerate}
    \end{tcolorbox}
    \caption{Recommendations for educators in addressing AI.}
    \label{fig-rec-updated}
\end{figure}

\subsection{Concluding Thoughts}
AI is set to change how we approach education. Drawing from the findings of this study and the broader literature, it's clear that within the realm of Physics education, AI presents a spectrum of threats and opportunities that vary based on context. Assessments at earlier educational stages, such as GCSE and A-Level, are notably susceptible when they are open-book. In contrast, when addressing advanced topics - especially at the university level and in textbook work - AI does not consistently provide correct answers, regardless of the prompting style. Moreover, students producing a high volume of quality work should not be unwelcome. The primary concern should be the active and meaningful involvement of students in creating such work. The path ahead remains uncertain; forthcoming Foundation models might bring about marginal enhancements or represent substantial breakthroughs in capabilities. With sustained research, assessment, and collaboration, the academic community has the opportunity to channel the potential of AI, ensuring it enhances, rather than diminishes, Physics education.
\end{section}

\newpage
\appendix
\section{AI Marking Prompt}
\begin{figure}[htp]
    \centering
    \begin{verbatim}
""" 
Your task is to assess an answer based on the provided format:
    Question: []
    Marks: []
    Solution: []     
    Guidance: []
    Answer: []

Square brackets contain necessary information. Based on the question,
solution, and any guidance, assess the answer's correctness. Return
only a number indicating the marks.

Example:
Question: [Interference fringes, produced by monochromatic light...]
Marks: [1]
Solution: [C]
Guidance: []
Answer: [C]

Correct Response: 1
Incorrect Response: This answer is correct.

Always return a number from 0 to 9. Responses are tested using Python
'.isdigit()' method. Any non-numeric answer will be sent back for re-
evaluation.

To assess:
Question: [{question}]
Marks: [{marks}]
Solution: [{solution}]     
Guidance: [{guidance}]
Answer: [{answer}] 
"""
    \end{verbatim}
    \caption{Condensed system prompt for AI self-marking. The AI was programmed to return a numerical score based on the question's solution and guidance. The full prompt, with multiple detailed examples, is abbreviated here for brevity. The AI accurately marked questions 58.8\% of the time.}
    \label{fig-marking}
\end{figure}

\FloatBarrier
\printbibliography

@article{Yeadon_2023,
  doi = {10.1088/1361-6552/acc5cf},
  url = {https://dx.doi.org/10.1088/1361-6552/acc5cf},
  year = {2023},
  month = {4},
  publisher = {IOP Publishing},
  volume = {58},
  number = {3},
  pages = {035027},
  author = {Will Yeadon and Oto-Obong Inyang and Arin Mizouri and Alex Peach and Craig P Testrow},
  title = {The death of the short-form physics essay in the coming AI revolution},
  journal = {Physics Education}
}

@article{transformers,
  title={Attention is all you need},
  author={Vaswani, Ashish and Shazeer, Noam and Parmar, Niki and Uszkoreit, Jakob and Jones, Llion and Gomez, Aidan N and Kaiser, {\L}ukasz and Polosukhin, Illia},
  journal={Advances in neural information processing systems},
  volume={30},
  year={2017}
}

@article{prior1,
  title={Education in the era of generative artificial intelligence (AI): Understanding the potential benefits of ChatGPT in promoting teaching and learning},
  author={Baidoo-Anu, David and Owusu Ansah, Leticia},
  journal={Available at SSRN 4337484},
  year={2023}
}

@article{prior2,
  title={ChatGPT: Bullshit spewer or the end of traditional assessments in higher education?},
  author={Rudolph, J{\"u}rgen and Tan, Samson and Tan, Shannon},
  journal={Journal of Applied Learning and Teaching},
  volume={6},
  number={1},
  year={2023}
}

@misc{prior3,
  title={Artificial intelligence in teaching and learning: what questions should we ask of ChatGPT?},
  author={Rospigliosi, Pericles ‘asher’},
  journal={Interactive Learning Environments},
  volume={31},
  number={1},
  pages={1--3},
  year={2023},
  publisher={Taylor \& Francis}
}

@article{chatBotComparison,
  title={War of the chatbots: Bard, Bing Chat, ChatGPT, Ernie and beyond. The new AI gold rush and its impact on higher education},
  author={Rudolph, J{\"u}rgen and Tan, Shannon and Tan, Samson},
  journal={Journal of Applied Learning and Teaching},
  volume={6},
  number={1},
  year={2023}
}

@article{yeadon2023exploring,
  title={Exploring Durham University Physics exams with Large Language Models},
  author={Yeadon, Will and Halliday, Douglas P},
  journal={arXiv preprint arXiv:2306.15609},
  year={2023}
}

@inproceedings{CoT,
 author = {Wei, Jason and Wang, Xuezhi and Schuurmans, Dale and Bosma, Maarten and ichter, brian and Xia, Fei and Chi, Ed and Le, Quoc V and Zhou, Denny},
 booktitle = {Advances in Neural Information Processing Systems},
 editor = {S. Koyejo and S. Mohamed and A. Agarwal and D. Belgrave and K. Cho and A. Oh},
 pages = {24824--24837},
 publisher = {Curran Associates, Inc.},
 title = {Chain-of-Thought Prompting Elicits Reasoning in Large Language Models},
 url = {https://proceedings.neurips.cc/paper_files/paper/2022/file/9d5609613524ecf4f15af0f7b31abca4-Paper-Conference.pdf},
 volume = {35},
 year = {2022}
}

@inproceedings{ZS-CoT,
 author = {Kojima, Takeshi and Gu, Shixiang (Shane) and Reid, Machel and Matsuo, Yutaka and Iwasawa, Yusuke},
 booktitle = {Advances in Neural Information Processing Systems},
 editor = {S. Koyejo and S. Mohamed and A. Agarwal and D. Belgrave and K. Cho and A. Oh},
 pages = {22199--22213},
 publisher = {Curran Associates, Inc.},
 title = {Large Language Models are Zero-Shot Reasoners},
 url = {https://proceedings.neurips.cc/paper_files/paper/2022/file/8bb0d291acd4acf06ef112099c16f326-Paper-Conference.pdf},
 volume = {35},
 year = {2022}
}

@inproceedings{SC-CoT,
title	= {Self-Consistency Improves Chain of Thought Reasoning in Language Models},
author	= {Xuezhi Wang and Jason Wei and Dale Schuurmans and Quoc V. Le and Ed H. Chi and Sharan Narang and Aakanksha Chowdhery and Denny Zhou},
year	= {2023},
URL	= {https://arxiv.org/abs/2203.11171},
booktitle	= {ICLR 2023}
}

@article{bitzenbauer2023chatgpt,
  title={ChatGPT in physics education: A pilot study on easy-to-implement activities},
  author={Bitzenbauer, Philipp},
  journal={Contemporary Educational Technology},
  volume={15},
  number={3},
  pages={ep430},
  year={2023},
  publisher={Bastas}
}

@article{tian2023chatgpt,
  title={Is ChatGPT the Ultimate Programming Assistant--How far is it?},
  author={Tian, Haoye and Lu, Weiqi and Li, Tsz On and Tang, Xunzhu and Cheung, Shing-Chi and Klein, Jacques and Bissyand{\'e}, Tegawend{\'e} F},
  journal={arXiv preprint arXiv:2304.11938},
  year={2023}
}

@article{dahlkemper2023physics,
  title={How do physics students evaluate artificial intelligence responses on comprehension questions? A study on the perceived scientific accuracy and linguistic quality of ChatGPT},
  author={Dahlkemper, Merten Nikolay and Lahme, Simon Zacharias and Klein, Pascal},
  journal={Physical Review Physics Education Research},
  volume={19},
  number={1},
  pages={010142},
  year={2023},
  publisher={APS}
}

@article{gregorcic2023chatgpt,
  title={ChatGPT and the frustrated Socrates},
  author={Gregorcic, Bor and Pendrill, Ann-Marie},
  journal={Physics Education},
  volume={58},
  number={3},
  pages={035021},
  year={2023},
  publisher={IOP Publishing}
}

@article{cooper2023examining,
  title={Examining science education in chatgpt: An exploratory study of generative artificial intelligence},
  author={Cooper, Grant},
  journal={Journal of Science Education and Technology},
  volume={32},
  number={3},
  pages={444--452},
  year={2023},
  publisher={Springer}
}

@article{kohnke2023chatgpt,
  title={ChatGPT for language teaching and learning},
  author={Kohnke, Lucas and Moorhouse, Benjamin Luke and Zou, Di},
  journal={RELC Journal},
  pages={00336882231162868},
  year={2023},
  publisher={SAGE Publications Sage UK: London, England}
}

@article{adiguzel2023revolutionizing,
  title={Revolutionizing education with AI: Exploring the transformative potential of ChatGPT},
  author={Adiguzel, Tufan and Kaya, Mehmet Haldun and Cansu, Fatih K{\"u}r{\c{s}}at},
  journal={Contemporary Educational Technology},
  volume={15},
  number={3},
  pages={ep429},
  year={2023},
  publisher={Bastas}
}

@article{reflection,
  title={Reflexion: an autonomous agent with dynamic memory and self-reflection},
  author={Shinn, Noah and Labash, Beck and Gopinath, Ashwin},
  journal={arXiv preprint arXiv:2303.11366},
  year={2023}
}

@article{chatGPTmathCapabilities,
  title={Mathematical capabilities of chatgpt},
  author={Frieder, Simon and Pinchetti, Luca and Griffiths, Ryan-Rhys and Salvatori, Tommaso and Lukasiewicz, Thomas and Petersen, Philipp Christian and Chevalier, Alexis and Berner, Julius},
  journal={arXiv preprint arXiv:2301.13867},
  year={2023}
}

@inproceedings{fewShot,
  author = {Brown, Tom and Mann, Benjamin and Ryder, Nick and Subbiah, Melanie and Kaplan, Jared D and Dhariwal, Prafulla and Neelakantan, Arvind and Shyam, Pranav and Sastry, Girish and Askell, Amanda and Agarwal, Sandhini and Herbert-Voss, Ariel and Krueger, Gretchen and Henighan, Tom and Child, Rewon and Ramesh, Aditya and Ziegler, Daniel and Wu, Jeffrey and Winter, Clemens and Hesse, Chris and Chen, Mark and Sigler, Eric and Litwin, Mateusz and Gray, Scott and Chess, Benjamin and Clark, Jack and Berner, Christopher and McCandlish, Sam and Radford, Alec and Sutskever, Ilya and Amodei, Dario},
  booktitle = {Advances in Neural Information Processing Systems},
  editor = {H. Larochelle and M. Ranzato and R. Hadsell and M.F. Balcan and H. Lin},
  pages = {1877--1901},
  publisher = {Curran Associates, Inc.},
  title = {Language Models are Few-Shot Learners},
  url = {https://proceedings.neurips.cc/paper_files/paper/2020/file/1457c0d6bfcb4967418bfb8ac142f64a-Paper.pdf},
  volume = {33},
  year = {2020}
}

@misc{openai_best,
  title = {Best practices for prompt engineering with OpenAI API},
  author = {OpenAI},
  year = {2023},
  month = {06},
  note = {\url{https://help.openai.com/en/articles/6654000-best-practices-for-prompt-engineering-with-openai-api}}
}

@article{dong2022survey,
  title={A survey for in-context learning},
  author={Dong, Qingxiu and Li, Lei and Dai, Damai and Zheng, Ce and Wu, Zhiyong and Chang, Baobao and Sun, Xu and Xu, Jingjing and Sui, Zhifang},
  journal={arXiv preprint arXiv:2301.00234},
  year={2022}
}

@online{openai_api,
  title={OpenAI GPT-3 Models},
  author={OpenAI},
  year={2020},
  url={https://platform.openai.com/docs/guides/gpt}
}

@online{openai_temperature,
  title={OpenAI API Documentation: Parameters},
  author={OpenAI},
  year={2020},
  url={https://platform.openai.com/docs/api-reference/completions/create}
}

@article{gptBiasEnglish,
title = {GPT detectors are biased against non-native English writers},
journal = {Patterns},
volume = {4},
number = {7},
pages = {100779},
year = {2023},
issn = {2666-3899},
doi = {https://doi.org/10.1016/j.patter.2023.100779},
url = {https://www.sciencedirect.com/science/article/pii/S2666389923001307},
author = {Weixin Liang and Mert Yuksekgonul and Yining Mao and Eric Wu and James Zou}
}

@article{promptEngineer1,
  title={Prompt Engineering with ChatGPT: A Guide for Academic Writers},
  author={Giray, Louie},
  journal={Annals of Biomedical Engineering},
  pages={1--5},
  year={2023},
  publisher={Springer}
}

@article{promptEngineer2,
  title={The CLEAR path: A framework for enhancing information literacy through prompt engineering},
  author={Lo, Leo S},
  journal={The Journal of Academic Librarianship},
  volume={49},
  number={4},
  pages={102720},
  year={2023},
  publisher={Elsevier}
}

@article{codePerform1,
  title={CodeGen: An Open Large Language Model for Code with Multi-Turn Program Synthesis},
  author={Nijkamp, Erik and Pang, Bo and Hayashi, Hiroaki and Tu, Lifu and Wang, Huan and Zhou, Yingbo and Savarese, Silvio and Xiong, Caiming},
  journal={arXiv e-prints},
  pages={arXiv--2203},
  year={2022}
}

@article{codePerform2,
  title={Evaluating large language models trained on code},
  author={Chen, Mark and Tworek, Jerry and Jun, Heewoo and Yuan, Qiming and Pinto, Henrique Ponde de Oliveira and Kaplan, Jared and Edwards, Harri and Burda, Yuri and Joseph, Nicholas and Brockman, Greg and others},
  journal={arXiv preprint arXiv:2107.03374},
  year={2021}
}

@article{detect1,
  title={Check Me If You Can: Detecting ChatGPT-Generated Academic Writing using CheckGPT},
  author={Liu, Zeyan and Yao, Zijun and Li, Fengjun and Luo, Bo},
  journal={arXiv preprint arXiv:2306.05524},
  year={2023}
}

@article{detect2,
  title={Game of Tones: Faculty detection of GPT-4 generated content in university assessments},
  author={Perkins, Mike and Roe, Jasper and Postma, Darius and McGaughran, James and Hickerson, Don},
  journal={arXiv preprint arXiv:2305.18081},
  year={2023}
}

@article{injection,
  title={More than you've asked for: A Comprehensive Analysis of Novel Prompt Injection Threats to Application-Integrated Large Language Models},
  author={Greshake, Kai and Abdelnabi, Sahar and Mishra, Shailesh and Endres, Christoph and Holz, Thorsten and Fritz, Mario},
  journal={arXiv preprint arXiv:2302.12173},
  year={2023}
}
\end{document}